\documentclass[a4paper]{jpconf}
\usepackage{graphicx}
\begin{document}
\title{ALICE ITS upgrade for LHC Run 3: commissioning in the laboratory}

\author{\textsc{Domenico Colella}$^{1}$ for the ALICE Collaboration}

\address{$^{1}$ Politecnico and INFN, Bari }

\ead{domenico.colella@ba.infn.it}

\begin{abstract}
ALICE is the CERN LHC experiment optimised for the study of the strongly interacting matter produced in heavy-ion collisions and devoted to the characterisation of the quark-gluon plasma. To achieve the physics program for LHC Run 3, a major upgrade of the experimental apparatus is ongoing. A key element of the upgrade is the substitution of the Inner Tracking System (ITS) with a completely new silicon-based detector whose features will allow the reconstruction of rare physics channels, not accessible with the previous layout. The enabling technology for such a performance boost is the adoption of custom-designed CMOS MAPS as detecting elements.
The installation of the detector in the ALICE cavern is completed and few months of commissioning will allow the final integration into the central systems.
In this talk, an overview of the adopted technologies as well as the status of the detector commissioning will be given.
\end{abstract}

\section{ALICE experiment upgrade and Inner Tracking System during Run 3}
ALICE (A Large Ion Collider Experiment) \cite{JINST} is a general-purpose, heavy-ion experiment at the CERN LHC. Its main goal is to study the physics properties of the quark-gluon plasma. 
In LHC Run 3 and Run 4  ALICE will focus on rare processes such as production of heavy-flavour particles, quarkonium states, real and virtual photons and heavy nuclear states \cite{ALICEupLoI}. 
The earlier methods of triggering will be limited for many of these measurements, particularly at low-$p_{\rm T}$. Therefore, the ALICE collaboration planned to upgrade the Run 1/Run 2 detector by enhancing its low-momentum vertexing and tracking capability, allowing data taking at substantially higher rates and preserving the already remarkable particle identification capabilities.
The upgraded experimental apparatus is designed to readout all \mbox{Pb--Pb} interactions, accumulating events corresponding to an integrated luminosity of more than 10 $nb^{-1}$. This minimum-bias data sample will provide an increase in statistics by about a factor 100 with respect to data collected during Run 1/Run 2 data taking. 

The main goals of the ITS upgrade (ITS 2) are an improved reconstruction of the primary vertex as well as decay vertices originating from heavy-flavour hadrons, and an enhanced performance for the detection of low-$p_{\rm T}$ particles. 
The design objectives are to improve the impact-parameter resolution by a factor of 3 and 5 in the r$\phi$- and z-coordinate, respectively, at a $p_{\rm T}$ of \mbox{500 MeV/c} \cite{ITSUPTDR}. The tracking efficiency and the $p_{\rm T}$ resolution at low $p_{\rm T}$ will also improve. Additio\-nal\-ly, the readout rate will be increased to \mbox{50 kHz} in \mbox{Pb--Pb} and \mbox{400 kHz} in \mbox{pp} collisions.
In order to achieve this performances, the following measures were taken:
\begin{itemize}
\item Granularity increased by an additional seventh layer and by equipping all layers with pixel sensors having cell size of  \mbox{29.24 $\mu$m $\times$ 26.88 $\mu$m}. 
\item Innermost detector layer moved closer to the interaction point, from \mbox{39 mm} to \mbox{22.4 mm}.
\item Material budget reduced to 0.3\% X$_{0}$ per layer for the innermost layers and to 1.0 \% X$_{0}$ for the outer layers.
\end{itemize}

The ITS 2 will be the first large-area silicon tracker based on the CMOS Monolithic Active Pixel Sensor (MAPS) technology operating at a collider. The basic element of a layer is the stave, which consists of a carbon space frame, supporting the chips, to which the cold plate and the cooling ducts are attached. Based on geometrical position and a few mechanical characteristics, the detector can be pictured as assembled in two parts: the innermost three layers form the Inner Barrel (IB); the middle two and the outermost two layers form the Outer Barrel (OB). Layer radial positions were optimised to achieve the best combined performance in terms of pointing resolution, $p_{\rm T}$ resolution and tracking efficiency in the expected high track-density environment of a \mbox{Pb--Pb} collision. The detector extends over a total surface of 10.3 m$^{2}$ and contains about \mbox{12.5 $\times$ 10$^{9}$} pixels with binary readout.
The ALPIDE, the pixel chip developed for the ITS 2, has been designed to fulfil the requirements of both the IB and OB. It is manufactured using the TowerJazz \mbox{180 nm} CMOS Imaging Sensor Process, has dimensions \mbox{15 $\times$ 30 mm$^{2}$} and hosts \mbox{512 $\times$ 1024} pixels.
Each pixel cell contains a sensing diode, a front-end amplifier and shaping stage, a discriminator, as well as a multi-hit buffer. The pixels are arranged in double columns and read out by a priority encoder which sends the addresses of the pixels that recorded a hit to the chip peripheral readout circuitry. 
The digital periphery produces the digitized and zero-suppressed hit data and sends them to the Readout Units (RUs) through high-speed serial data links running at either \mbox{1.2 Gbps} for the IB sensors or \mbox{400 Mbps} for the OB sensors. Chip periphery interfaces also with a clock input, and a bi-directional control bus for configuration, control, and monitoring. 
The RUs, FPGA-based boards, are located outside the detector acceptance in a moderate radiation environment. They receive the triggers from the Central Trigger Processor (CTP) via optical link and ship the data to the Common Readout Units housed in the First Level Processor computers, located in control rooms, where the event reconstruction starts. The same data path is used to control the detector. Power to the stave is provided through a Power Board (PB), generating two \mbox{1.8 V} supply voltages (for the analog and digital circuits of the pixel chip) and a negative voltage output for the reverse bias. The RUs and PUs are powered by a CAEN distribution system based on EASY3000 crates. 

\section{Commissioning in the laboratory}
Detector was assembled in half-barrels (two, top and bottom, of the IB and two, top and bottom, of the OB) in a large clean room at CERN equipped with the same backend system that will be used in the experiment, including powering system, cooling system, full readout and trigger chains. The first completely assembled and connected half-barrel has been the IB top by July 2019; last one has been the OB bottom by January 2020. 
Continuous operation of the detector started in May 2019, supported by shift crews alternating along the day every 8 hours. Most of the activities of the standard shift crews have been performed on the IB. Standard data taking schedule foresees hourly execution of a threshold scan, a fake-hit rate run  and a readout test. OB took longer for the final assembly and required more time for the complete integration in the system, due to larger number of elements. Most of the verifications performed on the OB staves have been done by detector experts.

\subsection{Inner Barrel}
\begin{figure}[t]
\centering
\includegraphics[width=37pc,clip]{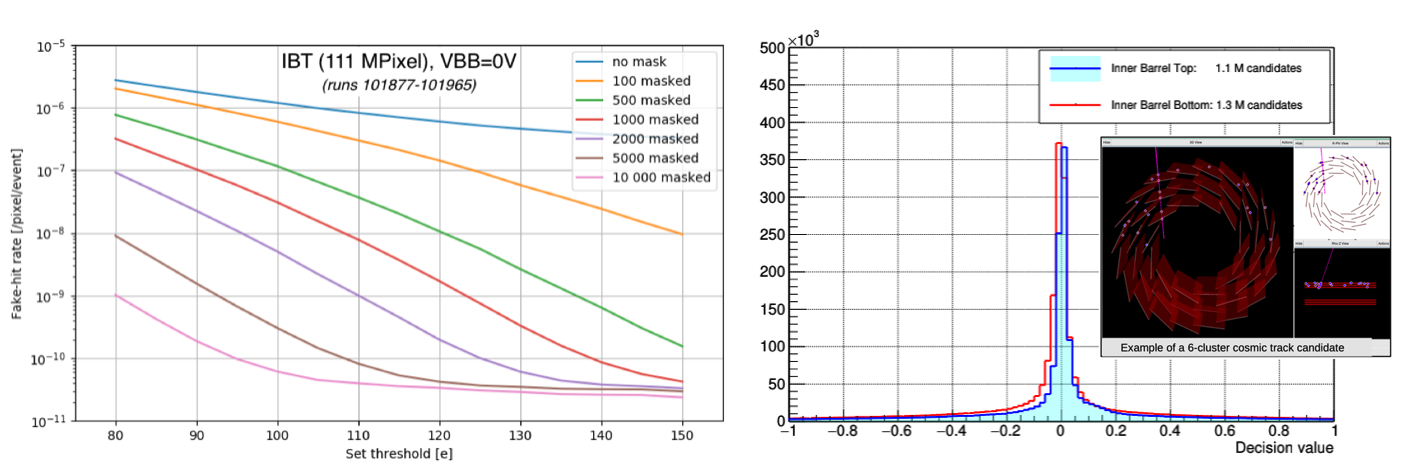}
\caption{(Left) Fake-hit rate as a function of the threshold value for different numbers of masked pixels. Results obtained for the full IB top. (Right) Cluster correlation distribution from cosmics tracks. In the event display, an example of tracks reconstructed from 6 aligned hits, obtained thanks to the overlap of the chips at the edges.}
\label{fig-4}
\end{figure}

During fake-hit rate runs, random triggers are sent to the chips and all the hits created in coincidence are collected. These hits come from the noise and also from cosmic rays crossing the detector in coincidence with the trigger. In the left plot of Fig. \ref{fig-4}, the measured fake-hit rate is plotted as a function of the threshold value for different numbers of masked noisier pixels. It can be seen that, for the target threshold value of \mbox{100 e$^{-}$}, masking a total of 10000 pixels, less than the 0.009\% of the considered acceptance, a fake-hit rate value of the order of \mbox{10$^{-10}$ hits/pixel/event} can be achieved. 
This can even be improved applying a back-bias voltage of -3 V to the chips, reaching a fraction of masked pixels of the order of 0.0001\%.
This is an important result, considering the really low fraction of masked acceptance and the fact that the target value for the fake-hit rate was \mbox{10$^{-6}$ hits/pixel/event} \cite{ITSUPTDR}. 

Due to the very low recorded chip noise, hits coming from cosmic particles passing through the detector can be clustered together and used to reconstruct the associated tracks. Starting from tracks one can perform an alignment study to determine the displacement of staves in layers obtaining residuals of the order of \mbox{100 $\mu$m}. The overlap of the staves at the edges facilitates the reconstruction of a track from 6 aligned clusters in three layers (right plot of Fig. \ref{fig-4}).

The quality selection criterion to consider a stave good for installation in the IB was to have a total amount of not working pixels below 50 among the 9 chips. After installation, the total number of pixels found to be dead were less than 1000 out of a total \mbox{226 $\times$ 10$^{6}$} pixels of the full IB. This number is actually reduced with respect to the measurements done on the staves before installation in the barrel, thanks to improved stuck pixel identification and masking mechanism.   

Readout tests are used to verify the quality of the data transmission from the chips to the associated RU performing a measurement of the Bit Error Rate (BER) through counting of 8b10b decoding errors. During the test, a fixed number of pixels are stimulated and the chip matrix is read out at a given rate. Multiple injected patterns (256, 512, 1008, 2048 and 4096 pixels) and readout rates (\mbox{44.9 kHz}, \mbox{67.3 kHz}, \mbox{101 kHz}, \mbox{123 kHz}, \mbox{202 kHz} and \mbox{247 kHz}) were explored. From \cite{ITSUPTDR}, a total of $\sim$250 activated pixels are expected in case of \mbox{100 kHz} \mbox{Pb--Pb} minimum-bias collisions with the unavoidable QED background, adding pixel noise of the order of \mbox{10$^{-5}$ hits/pixel/event}, for an integration time of \mbox{30 $\mu$s}. Simulating an occupancy similar to the one described, a BER sensitivity of about 10$^{-12}$ was measured for all the readout rates mentioned above. The BER sensitivity was measured to be of the order of 10$^{-15}$ in a large statistics sample for the readout rate of 44.9 kHz. Under extreme conditions going significantly beyond the possible scenarios during data taking with beams, errors have been recorded on a limited amount of staves and with a rate of error of the order of $\sim$1/20 hours.

\subsection{Outer Barrel}
Handling of very long and fragile objects, as the staves during the assembly of the layers and half-barrel, required the definition of a rigorous and yet delicate procedure. In spite of the complexity of the stave installation procedure, there was no increase in the number of dead chips, as ascertained from electrical and functional checks. The total number of dead chips over the full OB is 32, corresponding to the 0.14\% of the acceptance; no overlap in the radial direction is present between the dead chips in the different layers. 

A calibration campaign allowed to characterise the OB staves, and showed that a tuning to \mbox{100 e$^{-}$} threshold target with a \mbox{2 e$^{-}$} precision is possible, having a chip-to-chip spread within the stave of \mbox{20 e$^{-}$}. 

\begin{figure}[t]
\centering
\includegraphics[width=37pc,clip]{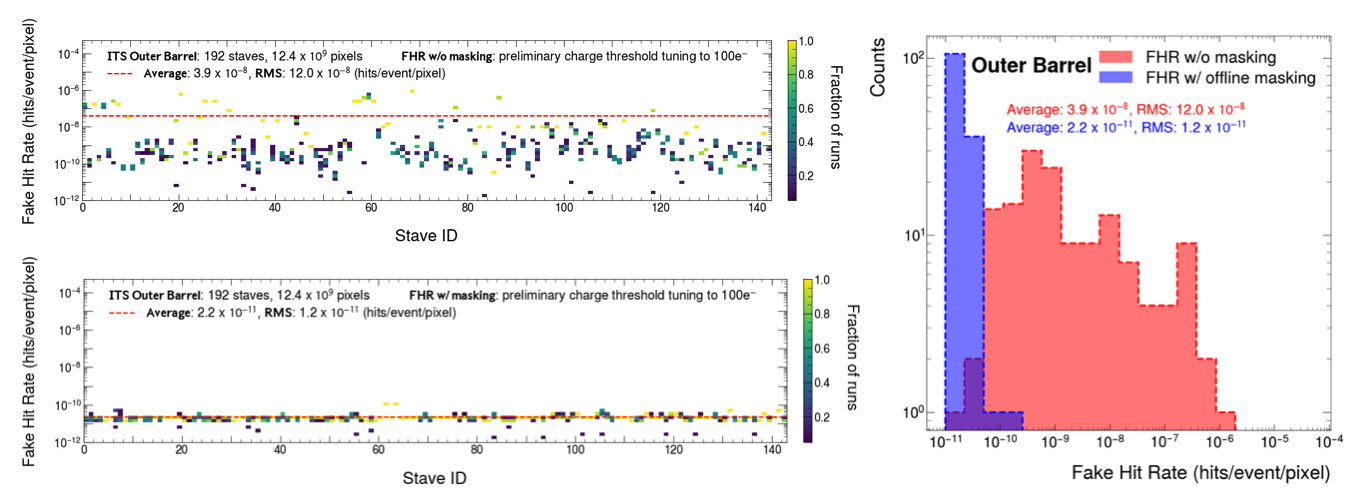}
\caption{First result of the OB stave fake-hit rate distribution without (top panel) and with (bottom panel) offline masking of noisier pixels.}
\label{fig-6}
\end{figure}

Due to the very large amount of chips installed in the OB staves ($\sim$24 $\times$ 10$^{3}$ chips), the selection criteria were released with respect to the chip used in the IB. As a consequence we do expect reduced performance in terms of noise. Nonetheless a fake-hit rate value of the order of \mbox{10$^{-11}$ hits/pixel/event} is reachable masking a mean value of  0.00036\% of the acceptance (Figure \ref{fig-6}).
This value is again well below the requirement of fake-hit rate to be less than \mbox{10$^{-6}$ hits/pixel/event} \cite{ITSUPTDR}.

A long term continuous powering campaign, to bring out events of early infant dead of staves and any service components (e.g. power supply modules), was carried out and no suspicious cases have been found. 

\section{Conclusions}
The ongoing replacement of the ITS will extend the physics reach of ALICE to lower transverse momentum, allowing the characterisation of the QGP via measurements of unprecedented precision. 
The new ITS is now placed within the ALICE detector in its final position. Few months of standalone and global commissioning will prepare the detector to be ready to take the very first data with collisions in the foreseen LHC pilot run during October 2021.


\end{document}